\documentclass[aps,prl,floatfix,twocolumn,showpacs]{revtex4}
\usepackage{amssymb}
\usepackage{amsmath}
\usepackage{graphicx}

\begin{document}

\title{Carbon Nanotubes as Cooper Pair Beam Splitters.}
\author{L.G. Herrmann$^{1,2,5}$, F. Portier$^{3}$, P. Roche$^{3}$, A. Levy Yeyati$^{4}$, T. Kontos$^{1,2}$\footnote{To whom correspondence should be addressed: kontos@lpa.ens.fr} and C. Strunk$^{5}$}
\affiliation{$^{1}$Ecole Normale Sup\'erieure, Laboratoire Pierre
Aigrain, 24, rue Lhomond, 75231 Paris Cedex 05,
France\\ $^{2}$CNRS UMR 8551, Laboratoire associ\'e aux universit\'es Pierre et Marie Curie et Denis Diderot, France\\
$^{3}$Service de physique de l'\'etat Condens\'e, CEA, 91192
Gif-sur-Yvette, France.\\
$^{4}$Departamento de F\'{\i}sica Te\'{o}rica de la Materia Condensada C-V, Universidad Aut\'onoma de Madrid, E-28049 Madrid, Spain.\\
$^{5}$ Institut f\"ur experimentelle und angewandte Physik ,
Universit\"at Regensburg, Universit\"atsstr.31,  93040 Regensburg,
Germany.}

\pacs{73.23.-b,73.63.Fg}

\begin{abstract}
We report on conductance measurements in carbon nanotube based
double quantum dots connected to two normal electrodes and a central
superconducting finger. By operating our devices as beam splitters,
we provide evidence for Crossed Andreev Reflections \textit{tunable
in situ}. This opens an avenue to more sophisticated quantum
optics-like experiments with spin entangled electrons.
\end{abstract}

\date{\today}
\maketitle

Quantum optics has been an important source of inspiration for many
recent experiments in nanoscale electric
circuits\cite{Feve:07,Wallraff:04}. One of the basic goals is the
generation of entangled electronic states in solid state systems.
Superconductors have been suggested as a natural source of spin
entanglement, due to the singlet pairing state of Cooper pairs. One
important building block required for the implementation of
entanglement experiments using superconductors is a Cooper pair beam
splitter which should split the singlet state into two different
electronic orbitals \cite{Martin:02,Buttiker:03}.

The basic mechanism for converting Cooper pairs into quasiparticles
is the Andreev reflection in which an originally quantum coherent
electron pair in the singlet spin state is produced at an interface
between a superconductor and a normal conductor. Conventional
Andreev Reflections (AR) are local and cannot readily be used to
create bipartite states\cite{Martin:96,Recher:01}. It has been
suggested to make use of electron-electron
interactions\cite{Recher:01,Martin:01,Choi:00,Recher:02,Bena:02,Bouchiat:03,Levy:07,Dupont:06},
spin filtering \cite{Feinberg:00} or anomalous scattering in
graphene \cite{Cayssol:08} to promote Cooper pair splitting i.e. the
Crossed Andreev Reflection (CAR) process.

\begin{figure}[!hpth]
\centering\includegraphics[height=1.0\linewidth,angle=0]{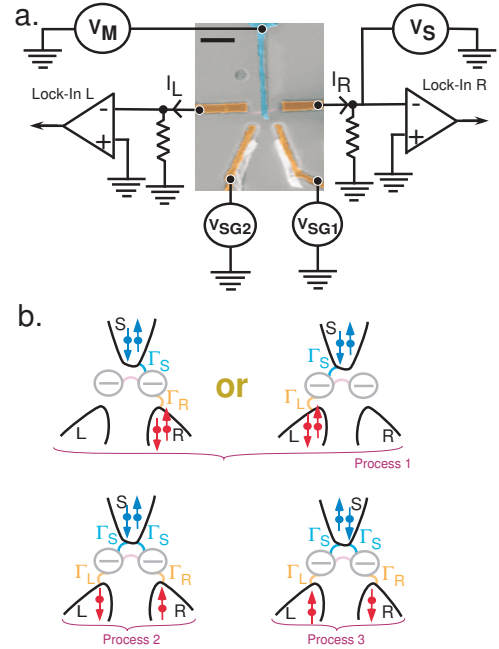}
\caption{ a. SEM image of a typical Cooper pair splitter device in
false colors with the two biasing schemes sketched. The bar is $1\mu
m$. A central superconducting electrode is connected to two quantum
dots engineered in the same single wall carbon nanotube (in purple)
which bridges between electrodes L and R. b. The elementary
processes which carry current in the superconducting (S) state. In
addition to the conventional local Andreev Reflection process, the
Crossed Andreev Reflection can occur in which a Cooper pair is split
in the two quantum dots.
 The relative probability of each of these processes can be inferred from the topology of the beam splitter.}%
\label{Figure1}%
\end{figure}

In this letter, we show that Coulomb interactions as well as size
quantization can favor the CAR processes in carbon nanotubes. We use
a double quantum dot geometry where the nanotube is connected to two
normal electrodes and a central superconducting finger. By operating
our device as a beam splitter (i.e. biasing the central
superconducting electrode), we find that there is a finite current
flowing from the superconducting electrode to the left (L) arm and
the right (R) arm for a bias voltage smaller than the energy gap of
the superconductor, which demonstrates Cooper pair injection. This
subgap current is enhanced when we tune the device to the degeneracy
points of the double dot with the help of capacitively coupled side
gate electrodes. This enhancement together with the dependence of
the asymmetry of transport in the superconducting state with respect
to the normal state provide evidence for Crossed Andreev Reflections
\textit{tunable in situ}, in contrast to the weakly interacting
metallic case\cite{Beckmann:04,Russo:05}. These conclusions are
supported by theoretical calculations based on a straightforward
modeling of the device.


We use chemical vapor deposition to fabricate our SWNTs, which are
localized with respect to Au alignment markers by Scanning Electron
Microscopy (SEM). A SEM picture of a typical device is shown in
figure \ref{Figure1}a. We fabricate our devices using standard
e-beam lithography and thin film deposition techniques. We deposit
the normal and the superconducting contact in one fabrication step
using shadow evaporation techniques. The normal contacts consist of
5nm of titanium followed by 50nm of Au or of 70nm of Pd. The central
superconducting electrode, $80nm-100nm$ wide, is a Al(100nm)/Pd(3nm)
bilayer. Such a method allows to achieve contact resistances as low
as 30kOhm between normal and superconducting reservoir. In addition
to the highly doped Si substrate with $500nm$ $SiO_{2}$ which is
used as a global backgate, we implement two side gates whose voltage
$V_{SG1}$ and $V_{SG2}$ can be tuned to control the two different
parts defined by the central superconducting electrode. The spacing
between the two normal contacts is between $600nm$ and $1.2\mu m$.
All the measurements presented in this letter have been carried out
in a dilution refrigerator with a base temperature of $80mK$, on one
particular sample which fulfilled the (stringent) requirements of
double dot spectroscopy and high enough coupling to the
superconducting electrode. The currents flowing through the
different arms of the beam splitter are measured  via the voltage
drop across two $2k\Omega$ resistors placed in series with the
device as shown on figure \ref{Figure1}a.

For characterization, we first operate the device as a series double
quantum dot by setting $V_M = 0$  and $V_S \neq 0$ using the bias
scheme shown in figure 1a. Figure 2a displays the color scale plot
of linear conductance $G_L = dI_L /dV_S$ through the right (R) arm
of the device as a function of side gate voltage 2 $V_{SG2}$ and
side gate voltage 1 $V_{SG1}$. The characteristic 'honeycomb'
stability diagram of a double quantum dot with rather regularly
spaced avoided crossings is observed.

\begin{figure}[!hpth]
\centering\includegraphics[height=0.85\linewidth,angle=0]{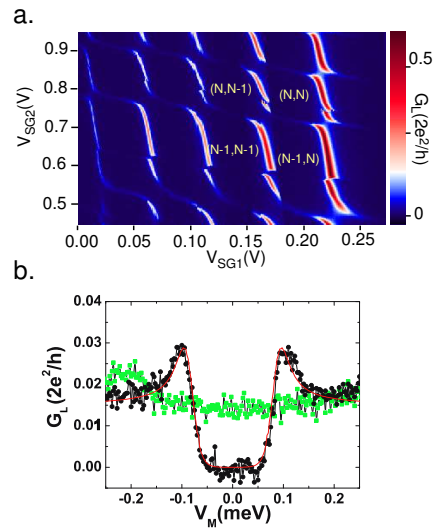}
\caption{a. Colorscale plot of the differential conductance $G_L$ as
a function of  side gate voltage 1 $V_{SG1}$ and side gate voltage 2
$V_{SG2}$. b. Differential conductance (black squares with lines)
measured out of resonance in the middle injection scheme at $100mK$
and $0mT$. In green square, the differential conductance at
$44.5mT$. In red solid lines, the BCS fit which yields an energy gap
$\Delta$
of $85 \mu eV$ and an electronic temperature of $100mK$. }%
\label{Figure2}%
\end{figure}

For the remaining of the letter, we operate the device as a beam
splitter by setting $V_{M} \neq 0$ and $V_{S}=0$. The differential
conductance corresponding to the left (L) arm and to the right (R)
arm of the beam splitter have qualitatively the same dependence as a
function of $V_{M}$, $V_{SG1}$ or $V_{SG2}$. If $V_{SG1}$ and
$V_{SG2}$ are tuned out of resonance, one typically measures a $G_L=
dI_L/dV_M$ shown in figure \ref{Figure2}b. This demonstrates
tunneling into a superconductor i.e. the current is strongly
suppressed for a bias voltage below the energy gap $\Delta$. Such a
feature has been used very recently to probe the quasiparticle
relaxation time in SWNTs\cite{Mason:09}. A fitting to the thermally
smeared BCS density of states gives $\Delta=85 \mu eV$ and an
electronic temperature of $100mK$. Therefore, for $V_{M}<85 \mu eV$,
one can only inject Cooper pairs. Note that the energy gap has a
reduced value with respect to pure Al since it corresponds to the
minigap of the Al/Pd bilayer \cite{Kontos:04}. As shown on figure
\ref{Figure2}b in green squares, the BCS gap does not appear if a
magnetic field of $44.5mT$ is applied perpendicularly to the axis of
the superconducting finger. This allows us to define the normal (N)
state of the beamsplitter for which we apply a field of $89mT$ in
order to be sure that superconductivity of the Al/Pd slab is absent.
The superconducting (S) state is obtained for $0mT$.

The colorscale plot of figure 3a displays $G_L$ for $V_M = 40 \mu
eV$ as a function of $V_{SG1}$ and $V_{SG2}$ for a specific
anti-crossing (AC1) in the S state. In contrast to the off resonance
case shown in figure 2b, a relatively high subgap current $I_S
\approx 0.1 I_N$ can flow into or from the superconductor even
though $V_M$ is smaller than the energy gap at AC1. The presence of
a subgap current can only be understood if Andreev reflections are
taking place at resonance. In order to characterize the type of
Andreev process occurring near the anticrossing, we show in figure
3b the variations of $G_L = dI_L/dV_M$ and $G_R = dI_R/dV_M$
measured simultaneously along the yellow arrow of the colorscale
plot in figure 3a. In the N state (black solid lines), $G_L$ and
$G_R$ display two peaks corresponding to the bonding/antibonding
states of the double dot. Due to an asymmetry of the coupling of
each dot to the L(R) reservoirs, the height of each doublet of peaks
is different. In the S state represented in red solid lines, the
peak height is reduced, but there is still a finite conductance for
the L(R) arm which follows essentially the resonances observed in
the N state. A similar feature is shown in figure 3c for another
anti-crossing (AC2) for which the asymmetry between $G_L$ and $G_R$
is slightly different. Note that the curves in the N state have been
scaled down by 1/3 for the sake of clarity.

\begin{figure}[!hpth]
\centering\includegraphics[height=1.15\linewidth,angle=0]{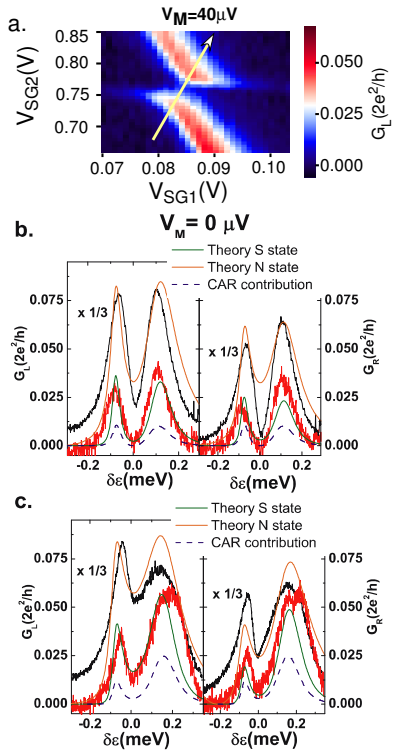}
\caption{a. Focus on a specific region for the differential
conductance $G_L$ as a function of $V_{SG1}$ and $V_{SG2}$ (AC1).
The yellow arrow indicates the direction in which the linear scan of
panel b. has been taken. b. Measurements of the conductance in the
normal state (black solid lines) and in the superconducting state
(red lines) for $G_R$ and $G_L$ along the direction of the yellow
arrow. For the sake of clarity, the N state conductances have been
multiplied by $1/3$. The model calculations are in green solid lines
for the S state and in orange solid lines for the N state. In blue
dashed lines, the CAR probability.
c. Similar graph as in b. for anticrossing 2 (AC2)}%
\label{Figure3}%
\end{figure}

The elementary processes which contribute to the subgap current in
our double quantum dot setup at the degeneracy point are
schematically sketched in figure 1b. The tunnel rates to the S,L and
R contacts are respectively  $\Gamma_S$, $\Gamma_L$ and  $
\Gamma_R$. The initial state in the superconductor S is represented
in blue and the final state in the normal metals N, in red. In
process 1, two particles are transferred to the same reservoir and
the corresponding probability is proportional to $\Gamma^{2}_{L,R}$.
In processes 2 and 3 which are equivalent, one particle is
transferred to each reservoir. The corresponding probability is
therefore proportional to $ \Gamma_L \Gamma_R$.  While the beam
splitter geometry imposes the above general form for the Andreev
tunneling probabilities, our theory (which we discuss in detail in
the EPAPS) allows to give an absolute value of the proportionality
constants of the AR and the CAR, by accounting for higher processes.

From the anti-crossing of figure 3a and the full stability diagram
of figure 2a, we determine all the important parameters of the
double dot using non-linear transport similarly to what is done in
ref. 20 for example. The corresponding plots obtained from our
theory at T = 0 are given in green solid lines for the S state and
in orange solid line for the N state for two anticrossings AC1 and
AC2. Each anticrossing is characterized by the set of parameters
$\{U_L, U_R,  \Gamma_{12}, \Gamma_{L}, \Gamma_{R}, \Gamma_{SL},
\Gamma_{SR}\}$, $U_{L,R}$ being the onsite charging energy,
$\Gamma_{12}$ the coupling between the two dots, $\Gamma_{L},
\Gamma_{R}$ the coupling of each dot to its reservoir and
$\Gamma_{SL}, \Gamma_{SR}$ the coupling of the superconductor to
each dot. Although not all of these parameters can be determined
independently, the fitting establishes the following hierarchy of
energy scales: $U_{L,R}\approx 1meV$, $\Gamma_{12}\approx 100 \mu
eV$, $\Gamma_{L}, \Gamma_{R} \approx 100 \mu eV$ and $\Gamma_{SL},
\Gamma_{SR} \approx 10 \mu eV$. For both arms and both
anticrossings, the calculated current in the superconducting state
is in good agreement with our experimental findings, which implies
an important contribution from CAR processes, as shown by the dashed
lines in figure 3b and 3c. For AC1, for example, the fit allows us
to extract a contribution of split Cooper pairs up to $35\%$ for
$G_L$ and $55\%$ for $G_R$.

\begin{figure}[!hpth]
\centering\includegraphics[height=0.85\linewidth,angle=0]{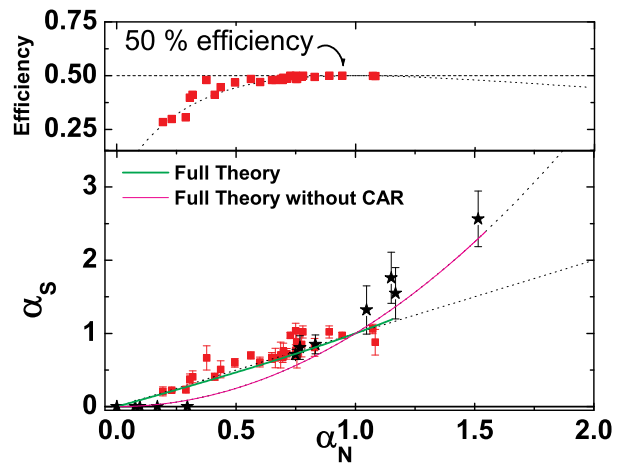}
\caption{Top panel : Efficiency of the Cooper pair splitting for the
device. The solid green line corresponds to the asymmetry variation
predicted by the interacting theory. Bottom panel : Asymmetry in the
superconducting state versus asymmetry in the normal state for 31
out of the 35 anti-crossings studied (red closed squares) and 11
resonance peaks away from the anti-crossings (black stars). The
dashed curve corresponds to $\alpha_S=\alpha^{2}_{N}$. The solid
dashed line corresponds to $\alpha_S=\alpha_{N}$. The green line is
the full
theory (see EPAPS) and the pink line is the full theory without the CAR process.}%
\label{Figure4}%
\end{figure}

 Now, we exploit the asymmetry of transport between the L and R
 arms to provide further evidence for the Cooper pair splitting. In the S state, it is
defined as $\alpha_S = G_L/G_R $ at 0mT,
 and in the N state, as $\alpha_N = G_L/G_R $ at 89mT, both taken at resonance. One can already anticipate
 from the elementary processes of figure 1b that the presence or the absence of processes 2 and 3
 should be observable directly from the dependence of $\alpha_S$ with respect to $\alpha_N$.
 At the degeneracy point, both the local and
 non-local Andreev processes acquire a Breit-Wigner-like form at resonance (see EPAPS). In the limit $\Gamma_{12}>>\Gamma_{L,R}>>\Gamma_{S}$, obtain for the conductance
$G_{L(R)}$:
\begin{equation}\label{eq:1}
G_{L(R)}=\frac{4e^2}{h}\frac{16
\widetilde{\Gamma}_{S}^{2}}{[\Gamma_{L}+\Gamma_{R}]^4}[\Gamma^{2}_{L(R)}+\Gamma_{L}\Gamma_{R}],
\end{equation}
where $\widetilde{\Gamma}_{S}$ is the tunnel rate to the
superconductor renormalized by Coulomb interactions. The presence of
the crossed term $\Gamma_L \Gamma_R$ in equation (1) implies that:
\begin{equation}\label{eq:2}
\alpha_{S}=\frac{\Gamma^{2}_{L}+\Gamma_{L}\Gamma_{R}}{\Gamma^{2}_{R}+\Gamma_{L}\Gamma_{R}}=\frac{\Gamma_{L}}{\Gamma_{R}}=\alpha_N.
\end{equation}
 Away from the degeneracy points, when one of the dots is blocked,
 only local AR are present and the term proportional to  $\Gamma_{L}\Gamma_{R}$ in equation (1) is absent.
 In this case, $\alpha_S$ is just the square of $\alpha_N$. The bottom panel of figure 4 displays $\alpha_S$ versus
 $\alpha_N$ for all the anti-crossings which we have measured, in red squares, and resonances
 away from the anticrossings, in black stars. The 31 red squares correspond to 7 different anticrossings.
 The error bars correspond to the systematic error made when determining $\alpha_S$. The black stars are obtained
 when one of the two dots is Coulomb blockaded (single resonance case).
 Within the error bars, they fall onto the dashed curve which corresponds to the universal
 parabola $\alpha_S=\alpha^{2}_{N}$ as expected. The red points fall onto the universal
 dashed line $\alpha_S=\alpha_{N}$ \cite{offpoints}. The observed contrast between the
 quadratic behavior followed by the black stars and the linear behaviour followed by the
 red squares proves the Cooper pair splitting action of our device. Note
that the full theory (i.e. without the approximations leading to Eq.
(1)) predicts a dependence of $\alpha_S$ versus $\alpha_{N}$ which
is marginally different from these behaviors  in our parameter
 range as shown by the green and pink curves in figure \ref{Figure4}.
 The universal linear behavior arising from the Cooper pair splitting
is therefore
 a very useful diagnosis tool for a wide range of parameters of our beam splitter. The efficiency of the
 beam splitter can be defined as the ratio $2T_{CAR}/(AR_{L} + AR_{R} + 2T_{CAR}) \approx 2/(2 + 1/\alpha_S + \alpha_S)$ of the split current
 to the total current. It is displayed on the top panel of figure \ref{Figure4}. For a $\alpha_N$ ranging from 0.5 to 1,
 the efficiency of Cooper pair splitting is close to $50\%$, showing
 that almost half of the Cooper pairs flowing out of the superconducting finger are split at the SWNT.

In conclusion, we have shown that carbon nanotube double quantum
dots can be used as tunable Cooper pair beam splitters. The specific
advantage of beam splitters based on double quantum dots is the
possibility of further processing the electron states e.g. by spin
filtering for EPR type experiments.

During the course of writing this paper, we became aware of a
similar study using InAs nanowires by Hofstetter et
al.\cite{Hofstetter:09}.

\begin{acknowledgments}
We thank A. Cottet for a critical reading of the manuscript and
illuminating discussions. We thank The Mesosocopics group of LPA,
The Quantronics Group and J. Siewert for assistance and fruitful
discussions. This work is supported by the SFB 689 of the Deutsche
Forschungsgemeinschaft, the ANR-05-NANO-028 contract, the
ANR-07-NANO-011-004 contract, the EU contract FP6-IST-021285-2, the
C'Nano Ile de France contract SPINMOL, the Spanish MICINN under
contract FIS2008-04209 and the DFH-UFA and DAAD mobility grants.
\end{acknowledgments}

\end{document}